\documentclass{WileyMSP-template}
\usepackage{xcolor}
\usepackage[version=4]{mhchem}
\begin{document}

\pagestyle{fancy}
\rhead{\includegraphics[width=2.5cm]{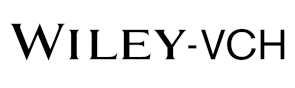}}

\title{Strain Effects in a Directly Bonded Diamond-on-Insulator Substrate}

\maketitle


\author{Ioannis Varveris*}
\author{Gianni D. Aliberti}
\author{Tianyin Chen}
\author{Filip A. Sfetcu}
\author{Diederik J. W. Dekker}
\author{Alfred Schuurmans}
\author{Nikolaj K. Nitzsche}
\author{Salahuddin Nur}
\author{Ryoichi Ishihara}



\begin{affiliations}
I. Varveris, G. D. Aliberti, F. Sfetcu, A. Schuurmans, N. K. Nitzsche, Prof. R. Ishihara\\
QuTech, Delft University of Technology, Lorentzweg 1, 2628 CJ, Delft, The Netherlands\\
Email: i.varveris@tudelft.nl

I. Varveris, G. D. Aliberti, T. Chen, D. J. W. Dekker, N. K. Nitzsche, Dr. S. Nur, Prof. R. Ishihara\\
Department of Quantum \& Computer Engineering, Delft University of Technology, Mekelweg 4, 2628 CD Delft, The Netherlands

\end{affiliations}


\keywords{Diamond-on-insulator (DOI), direct bonding, nitrogen-vacancy (NV) centers, optically detected magnetic resonance (ODMR), photoluminescence (PL) mapping, strain effects, thermal expansion mismatch}

\begin{abstract}
The direct bonding process of a diamond-on-insulator (DOI) substrate enables monolithic integration of diamond photonic structures for quantum computing by improving photon collection efficiency and entanglement generation rate between emitters. It also addresses key fabrication challenges, such as robustness, bonding strength, and scalability. In this study, we investigate strain effects in DOI substrates following direct bonding. Strain generation is expected near the diamond-SiO\textsubscript{2}/Si interface due to thermal expansion coefficient mismatch between the bonded materials. Strain-induced lattice distortions are characterized using nitrogen-vacancy (NV) centers in diamond via optically detected magnetic resonance (ODMR) and photoluminescence (PL) mapping. PL mapping reveals interference fringes in unbonded regions, indicating bonding irregularities. Depth-resolved ODMR measurements show a volumetric strain component increase of $\sim$0.45 MHz and a shear component increase of $\sim$0.71 MHz between the top surface and the DOI interface. However, ODMR signal contrast and peak linewidth remain largely unaffected, suggesting no visible deterioration in the optical properties of the emitters. By combining ODMR and PL mapping, this work establishes a robust methodology for assessing bonding quality and strain impact on NV centers, an essential step toward advancing scalable quantum technologies and integrated photonic circuits.
\end{abstract}


\section{Introduction}

One of the challenges in achieving robust quantum communication and high-fidelity quantum computing has long been the entanglement generation rate between NV centers, which is limited by the relatively low fraction of emitted coherent photons that can be detected. A solution to address this problem is to integrate NV centers into optical cavities with a high ratio of quality factor to mode volume. This is done to increase the zero-phonon line (ZPL) emission and improve collection efficiency \textsuperscript{[1]}. 

However, current applications are limited by the challenges of growing large single crystal diamond (SCD) via chemical vapor deposition (CVD), which restricts its use in integrated photonics \textsuperscript{[2]}. To overcome this, researchers are developing thin film diamond integrated with color centers in photonic crystal (PhC) cavities \textsuperscript{[3]}. Two prominent approaches are hybrid and monolithic all-diamond systems \textsuperscript{[4]}. Hybrid systems use non-diamond materials like silicon nitride or gallium phosphide coupled with diamond color centers for easier manufacturing and integration \textsuperscript{[5, 6, 7]}. Nevertheless, they face issues with limited coupling efficiency. Monolithic all-diamond systems, which entail fabricating optical cavities directly on single crystal diamond substrates, have shown success with not only NV \textsuperscript{[8]}, but also SiV \textsuperscript{[9, 10]}, and GeV centers \textsuperscript{[8]}. 

Such a monolithic fabrication technique is the direct bonding of diamond on insulator substrates. Hydrophilic (water-attracting) direct bonding, in particular, is preferred over indirect methods using adhesives, due to its high bonding strength, temperature stability, and industrial scalability \textsuperscript{[11]}. Similar to Silicon-On-Insulator (SOI) substrate bonding \textsuperscript{[12]}, Diamond-On-Insulator (DOI) bonding adheres to the same operating principles. After the surface of the diamond and the insulator (in this case SiO\textsubscript{2}) have been activated, they are then brought into contact, in order for them to create a bond, as a result of van der Waals forces, and form covalent bonds \textsuperscript{[12]}. The material is subsequently annealed in order to enhance the quality of the interface \textsuperscript{[13]}. 

However, despite the effectiveness of direct bonding and subsequent annealing in enhancing interface quality, this fabrication process can introduce unintended strain effects within the diamond structure. It is probable that the diamond, having undergone the aforementioned chemical processing, will exhibit further stress effects on the surface which is bonded to the insulator substrate. Furthermore, part of the direct bonding process takes place under a temperature of 200°C. During the cooling period, each of the bonded materials contracts at different rates, according to their unique thermal expansion coefficients. This mismatch between the thermal expansion rates between the diamond and the silicon dioxide/silicon can lead to the generation of strain at their interface \textsuperscript{[14, 15]}, and potentially further deep into the diamond lattice.

These strain effects can cause deformations to the crystal lattice, which result in inhomogeneous broadening and degradation of the spin dephasing time of the NV centers. Therefore, it is deemed critical to be able to understand and visually map the effects of strain, providing meaningful insight into the quality of the bonded samples.

\section{Direct Bonding Process Overview}

\begin{figure}[ht]
    \centering
    \includegraphics[width=\linewidth]{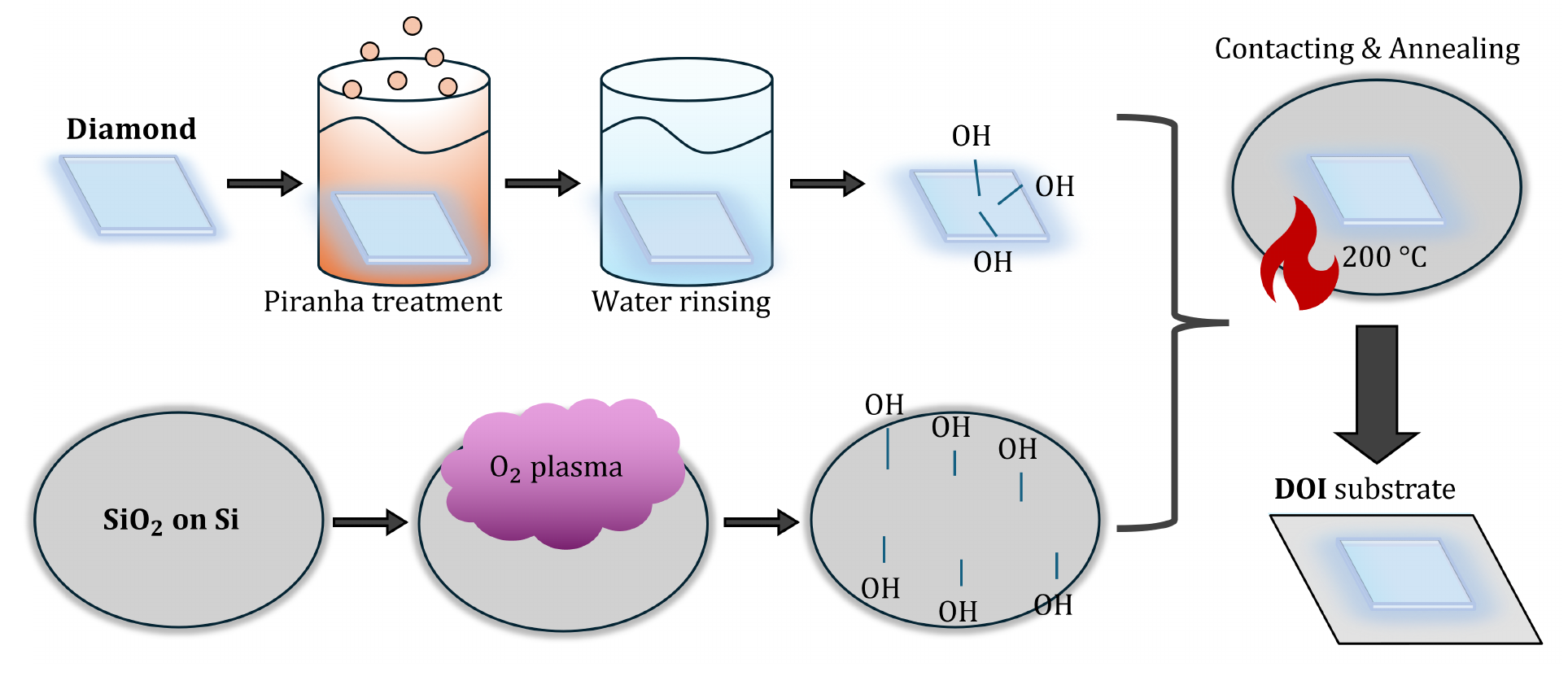}
    \caption{Schematic overview of the hydrophilic direct bonding process used to fabricate the diamond-on-insulator (DOI) substrate. The process begins with a type Ib, single-crystal, double-side polished, (100)-oriented diamond substrate, and a Si substrate, coated with a 300 nm PECVD SiO\textsubscript{2} layer, and diced into 25$\times$25 mm$^2$ chips. Both substrates undergo chemical activation. The diamond is cleaned using Piranha solution (3:1 H$_2$SO$_4$:H$_2$O$_2$) at 75$^\circ$C for 30 min and rinsed in deionized (DI) water for 5--10 min. Meanwhile, the SiO\textsubscript{2}/Si substrate is rinsed with DI water for 5 min, and is then subjected to O$_2$ plasma treatment (1000 W, 5 min, with 400 sccm O$_2$ flow) using a PVA Tepla system. These steps remove surface contaminants and enrich the surface with silanol (Si--OH) groups, promoting hydrophilicity. After cleaning, the substrates are brought into contact at room temperature (20$^\circ$C) and ambient humidity ($\sim$40\%), allowing van der Waals and hydrogen bonding to form across a thin interfacial water layer. This pre-bonding interaction is stabilized through a low-temperature anneal at 200$^\circ$C for 24 h, under N$_2$ flow. The annealing induces interfacial dehydration, resulting in covalent bond formation, thus completing the DOI heterostructure. The bonded stack typically exhibits high mechanical integrity with no visible voids or delamination.}
    \label{fig:bonding_schematic}
\end{figure}

The fabrication of the diamond-on-insulator (DOI) substrate utilized in this work follows a hydrophilic direct bonding process, outlined in Figure~\ref{fig:bonding_schematic}. No adhesives were used in the bonding, which relied entirely on surface chemical activation and low-temperature annealing to achieve interfacial adhesion.

Hydrophilic direct bonding is a process that involves creating a bond between surfaces treated to exhibit hydrophilic properties. The first step is surface activation, which prepares the surfaces for subsequent bonding by introducing reactive hydroxyl groups (–OH). For the diamond substrate, this involves thorough cleaning using a Piranha solution, a highly oxidative mixture of sulfuric acid (H\textsubscript{2}SO\textsubscript{4}) and hydrogen peroxide (H\textsubscript{2}O\textsubscript{2}) in a 3:1 ratio. The Piranha solution does not etch the diamond, but removes organic contamination and promotes surface hydrophilicity. This process results in the termination of the diamond surface with hydroxyl groups (C–OH), enhancing its reactivity, by providing essential reactive sites for bonding with other molecules and facilitating further chemical reactions \textsuperscript{[13, 16]}.

The insulator substrate is composed of a 300 nm, PECVD-grown layer of SiO\textsubscript{2} on Si. Hydrophilic bonding relies on the formation of silanol (Si–OH) groups on the SiO\textsubscript{2} surface. Therefore, following rinsing of the SiO\textsubscript{2}/Si substrate with DI water, it undergoes O\textsubscript{2} plasma treatment, a common method for cleaning and oxidizing silicon surfaces \textsuperscript{[13]}. This treatment both removes surface contamination and increases silanol group density by oxidizing siloxane (Si–O–Si) bridges in the presence of water vapor. These hydroxyl-terminated surfaces are hydrophilic and attract adsorbed water molecules from the environment.

\begin{figure}[ht]
    \centering
    \includegraphics[width=\linewidth]{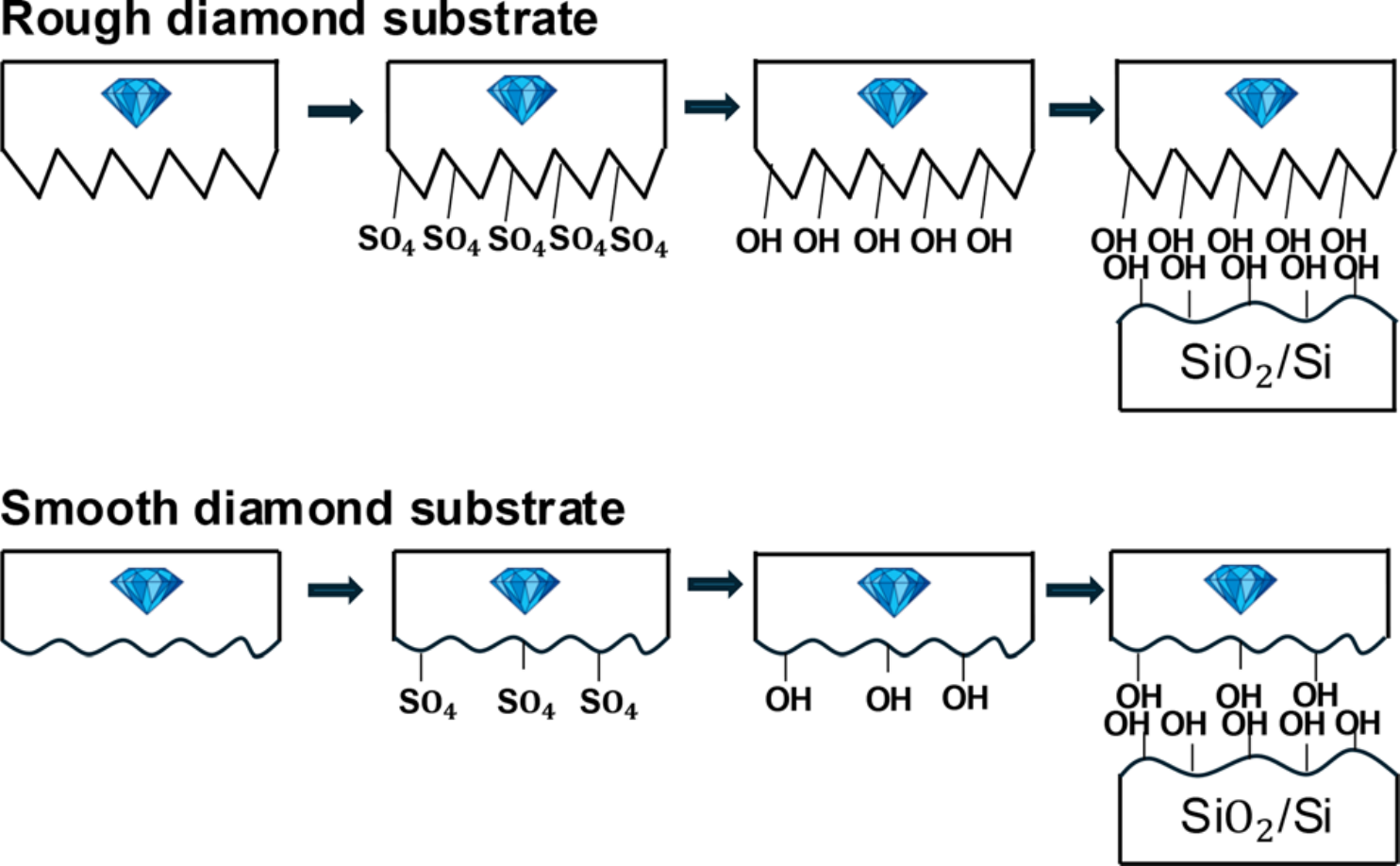}
    \caption{Illustration of the interfacial bonding mechanism between diamond and SiO\textsubscript{2}. Increased surface roughness enhances the density of hydroxyl (–OH) terminations, facilitating more effective hydrogen bonding and improving the overall bonding success. Reproduced with permission from Chen \textit{et al.}\textsuperscript{[11]}. © 2025 AIP Publishing.}
    \label{fig:bonding-mechanism}
\end{figure}

Once activated, both the diamond and SiO\textsubscript{2}/Si surfaces become highly hydrophilic due to the formation of surface hydroxyl (–OH) groups. This promotes the adsorption of ambient water molecules, leading to the formation of a thin hydration layer on each surface. When the substrates are brought into contact under ambient conditions, hydrogen bonding and van der Waals interactions across this interfacial water layer enable initial adhesion and stabilize the alignment of the surfaces (see Figure~\ref{fig:bonding-mechanism}). Specifically, hydroxyl groups such as C–OH and Si–OH participate in hydrogen bonding, which, although individually weak, collectively contribute significantly to the mechanical integrity of the interface\textsuperscript{[17]}.

Following initial contact, the DOI stack undergoes annealing, with temperature and duration serving as critical parameters for achieving robust interfacial bonding. Utilizing low-temperature annealing at 200$^\circ$C, the bond strength is enhanced through dehydration reactions between adjacent silanol groups, producing strong Si–O–C covalent bridges across the interface\textsuperscript{[17]}. This irreversible covalent bonding process completes the formation of the DOI heterostructure, as expressed by the reaction seen below:

\begin{center}
    \ce{C-OH + HO-Si -> C-O-Si + H2O}
\end{center}

Throughout this process, care was taken to maintain surface roughness within an optimal range. The  surface roughness values of the diamond and the SiO\textsubscript{2}/Si substrate fall within the 2–5 nm range typically reported as suitable for reliable hydrophilic bonding\textsuperscript{[11]}. Notably, no high-temperature processing or aggressive etching was applied that could significantly modify surface topography prior to bonding.

The bonded interface remains intact through post-processing steps, and no delamination or void formation is typically observed. This is consistent with the formation of strong interfacial bonds and the maintenance of a mechanically coherent interface. The process flow has been replicated multiple times, further validating its general applicability and reproducibility.


\section{Strain Effects in the Diamond-on-Insulator System}

\subsection{Origin and Mechanisms of Strain in Directly Bonded DOI Structures}

The direct bonding process of a diamond-on-insulator (DOI) substrate inevitably induces strain at the interface, with the most significant contribution arising from thermal expansion mismatch between the bonded materials\textsuperscript{[15]}, since prior to bonding, the residual stress in both the diamond and SiO\textsubscript{2}/Si substrates is minimal\textsuperscript{[18]}. The diamond used did not undergo post-growth annealing or epitaxy, which would otherwise introduce non-uniform stress\textsuperscript{[19]}. Additionally, optical inspection revealed no bowing, indicating negligible macroscopic stress. Similarly, the underlying substrate remains mechanically relaxed\textsuperscript{[18]}. Any pre-existing uniform residual stress is accounted for as a constant offset in the ODMR measurements, as the observed strain is induced post-bonding.

The annealing process at 200$^\circ$C, necessary for bond stabilization, introduces thermal strain in the DOI heterostructure upon cooling to room temperature. This strain arises due to the mismatch in thermal expansion coefficients between the constituent materials. Diamond exhibits a low thermal expansion coefficient, with $\alpha_{\mathrm{D}} \approx 1.1 \times 10^{-6}$ K$^{-1}$ at room temperature and $\alpha_{\mathrm{D}} \approx 2.3 \times 10^{-6}$ K$^{-1}$ at 200$^\circ$C\textsuperscript{[20]}. In contrast, amorphous silica (SiO\textsubscript{2}) has an expansion coefficient of $\alpha_{\mathrm{SiO_2}} \approx 2.5\times 10^{-6}$ K$^{-1}$ at room temperature\textsuperscript{[21]} and $\alpha_{\mathrm{SiO_2}} \approx 3.0\times 10^{-6}$ K$^{-1}$ at 200$^\circ$C\textsuperscript{[21, 22]}. Silicon, on the other hand, expands significantly more than the diamond, with $\alpha_{\mathrm{Si}} \sim 2.6 \times 10^{-6}$ K$^{-1}$ at room temperature and $\alpha_{\mathrm{Si}} \sim 3.3 \times 10^{-6}$ K$^{-1}$ at 200$^\circ$C\textsuperscript{[23, 24]}. Table~\ref{tab:CTE_values} summarizes the relevant thermal expansion coefficients. Since the SiO\textsubscript{2} layer is only 300 nm thick, while the underlying Si substrate is several hundred microns thick, the thermal and mechanical behavior of the bonded insulator stack is dominated by the silicon. The diamond layer, in turn, is constrained by this underlying Si substrate across the interfacial SiO\textsubscript{2} layer, which behaves elastically under these conditions. Upon cooling, the Si substrate contracts more than the diamond, imposing compressive in-plane strain on the diamond layer\textsuperscript{[25]}. For a temperature drop of $\Delta T \sim 180$ K, the resulting in-plane thermal mismatch strain is estimated as $\varepsilon \sim (\alpha_{\mathrm{Si}}^{200^\circ \mathrm{C}} - \alpha_{\mathrm{D}}^{200^\circ \mathrm{C}})\Delta T \approx -1.8 \times 10^{-4}$.

\begin{table}[ht]
    \centering
    \caption{Linear thermal expansion coefficients ($\alpha$) of materials in the DOI stack.}
    \label{tab:CTE_values}
    \renewcommand{\arraystretch}{1.3}
    \begin{tabular}{|c|c|c|}
    \hline
    \textbf{Material} & \textbf{$\alpha$ (RT)} ($\times 10^{-6}$ K$^{-1}$) & \textbf{$\alpha$ (200$^\circ$C)} ($\times 10^{-6}$ K$^{-1}$) \\
    \hline
    Diamond & $\sim 1.1$ & $\sim 2.3$ \\
    PECVD SiO$_2$ & $\sim 2.5$ & $\sim 3.0$ \\
    Silicon & $\sim 2.6$ & $\sim 3.3$ \\
    \hline
    \end{tabular}
\end{table}

While the SiO\textsubscript{2} layer serves as the direct interface between diamond and Si, it is important to consider whether it could act as a strain-relieving buffer. In particular, viscous relaxation or interfacial slippage could potentially mitigate strain transfer. However, previous studies report that viscous flow in SiO\textsubscript{2} requires temperatures above 900–950$^\circ$C\textsuperscript{[26]}, well beyond the 200$^\circ$C annealing conditions used here. At these lower temperatures, the oxide behaves elastically. Moreover, the formation of covalent interfacial bonds effectively suppresses slippage at the diamond–SiO\textsubscript{2} interface\textsuperscript{[11]}. As a result, the bonded stack behaves as a mechanically coherent system, with strain transmitted directly from the silicon substrate to the diamond through the thin, elastically deforming oxide layer.

Surface roughness is another potential contributor to strain. However, in our study, the bonded diamond surface had an initial roughness of $R_a < 2$ nm, preserved through the Piranha and DI water cleaning steps. O$_2$ plasma etching was applied to the unbonded top diamond surface, with no measurable change in surface roughness. The SiO\textsubscript{2} layer had a measured RMS roughness of $S_q = 4.44$ nm prior to bonding. Both these values fall within the 2--5 nm range considered suitable for uniform bonding\textsuperscript{[11]}.

Finally, localized surface defects, such as scratches or implantation damage, are known to produce sharp strain gradients and localized ODMR shifts\textsuperscript{[27]}. In contrast, the strain fields observed in our bonded samples are spatially extended and homogeneous over tens of microns, without any localized stress anomalies. Delamination is also not typically observed, following this direct bonding process. These features are inconsistent with surface-defect-induced strain, and instead match the expected thermomechanical strain distribution in a bonded multilayer stack. Therefore, the strain analyzed in this work originates predominantly from thermal expansion mismatch, transmitted through the SiO\textsubscript{2} layer without significant relaxation.

\subsection{Strain Physics in NV Centers}

The Hamiltonian $\hat{H}$ of the ground-state spin triplet state of the NV center in diamond, in the absence of imperfections (no strain) and external fields, can be described by the electron spin-spin interaction, which arises from second-order perturbation theory applied to the spin-orbit interaction \textsuperscript{[28]}. Defining the $z$-axis along the NV-axis results in the following expression for the Hamiltonian:

\begin{equation}
    \frac{\hat{H}}{h} = D S_z^2
\end{equation}
\\
where $h$ is Planck's constant, $\textbf{S}$ is the spin operator of the NV center, and $D=2.87$ GHz is the zero-field splitting (ZFS) tensor.

\subsubsection{Spin-Strain Interaction}

In reality, numerous imperfections exist within the diamond lattice, making the spin-strain interaction non-negligible. According to Udvarhelyi \textit{et al.} (2018) \textsuperscript{[29]}, the spin-strain interaction for the ground-state spin triplet can be described as a deformation in the diamond crystal, causing the strain tensor $\varepsilon_{ij}$ to couple to the NV spin through the symmetry-allowed Hamiltonian:

\begin{equation}
    \hat{H}_{\varepsilon} = \hat{H}_{\varepsilon}^{0} + \hat{H}_{\varepsilon}^{1} + \hat{H}_{\varepsilon}^{2}
\end{equation}
\\
where each term couples to different spin transitions. $\hat{H}_\varepsilon^0$ concerns the $\vert m_S=0\rangle$ to $\vert m_S=0\rangle$ transition (shifting or splitting the $\vert m_S=0 \rangle$ level), $\hat{H}_\varepsilon^1$ the $\vert m_S=0 \rangle$ to the $\vert m_S=\pm 1 \rangle$ transition, and $\hat{H}_\varepsilon^2$ the $\vert m_S=+1 \rangle$ to the $\vert m_S=-1 \rangle$ transition.

In the NV reference frame $(x,y,z)$, aligned so that $z$ is the NV-axis and $(x,y)$ spans the perpendicular plane, the general spin-strain Hamiltonian allowed by the $C_{3v}$ symmetry is written as \textsuperscript{[29]}:

\begin{equation}
\begin{array}{rl}
    \hat{H}_\varepsilon /h = &
 \Big\{ 
[h_{41}(\varepsilon_{xx} + \varepsilon_{yy}) + h_{43}\,\varepsilon_{zz}] S_z^2  \\[6pt]
&+ \frac{1}{2}[h_{26}\,\varepsilon_{zx} - \frac{1}{2}h_{25}(\varepsilon_{xx} - \varepsilon_{yy})]\{S_x,S_z\} \\[6pt]
&+ \frac{1}{2}[h_{26}\,\varepsilon_{yz} + h_{25}\,\varepsilon_{xy}]\{S_y,S_z\} \\[6pt]
&+ \frac{1}{2}[h_{16}\varepsilon_{zx} - \frac{1}{2}h_{15}(\varepsilon_{xx}- \varepsilon_{yy})](S_y^2 - S_x^2) \\[6pt]
&+ \frac{1}{2}[h_{16}\,\varepsilon_{yz} + h_{15}\,\varepsilon_{xy}]\{S_x,S_y\} \Big\} \\[6pt]
& \equiv M_z S_z^2  + N_x\{S_x,S_z\} + N_y\{S_y,S_z\} + M_x (S_y^2-S_x^2) + M_y\{S_x,S_y\}.
\end{array}
\label{Full_M-N_equation}
\end{equation}
\\
where $h_{ij}$ are coupling parameters (in units of GHz/strain) that depend on the microscopic electronic structure of the NV center (see Table~\ref{h_values}), $M_x$, $M_y$, and $M_z$ are the strain tensor amplitudes (in MHz) and $N_x$, $N_y$ are non-diagonal strain terms that induce coupling between $\vert m_S=0 \rangle$ and the $\vert m_S=\pm 1 \rangle$ states, and $\varepsilon_{ij}$ is the strain tensor, with $\mathbf{u}(\mathbf{r})$ being the displacement field:

\begin{equation}
\varepsilon_{ij} = \frac{1}{2} \left( \frac{\partial u_i}{\partial x_j} + \frac{\partial u_j}{\partial x_i} \right)
\end{equation}

If the resulting strain-induced mixing between $m_S=0$ and $m_S=\pm1$ states is small compared to the zero-field splitting $D$, the off-diagonal components $N_x$, $N_y$ may then be treated as a small perturbation and be neglected. 

\begin{table}[ht]
    \centering
    \caption{Spin-strain ($h$) coupling-strength parameters calculated from density functional theory by Udvarhelyi \textit{et al.} (2018) \textsuperscript{[29]}.}
    \label{tab:spin_strain}
    \renewcommand{\arraystretch}{1.3}
    \setlength{\tabcolsep}{10pt}
    \begin{tabular}{c c}
        \hline
        \hline
        \textbf{Parameter} & \textbf{Value (MHz/strain)} \\
        \hline
        $h_{43}$ & $2300 \pm 200$ \\
        $h_{41}$ & $-6420 \pm 90\phantom{0}$ \\
        $h_{25}$ & $-2600 \pm 80\phantom{0}$ \\
        $h_{26}$ & $-2830 \pm 70\phantom{0}$ \\
        $h_{15}$ & $5700 \pm 200$ \\
        $h_{16}$ & $19660 \pm 90\phantom{0}$ \\
        \hline
        \hline
    \end{tabular}
    \label{h_values}
\end{table}

\subsubsection{Eigenfrequencies Including the Zeeman Term}

In the presence of a homogeneous magnetic field $\mathit{B}$, the NV Hamiltonian acquires an extra term:

\begin{equation}
    \frac{\hat{H}_B}{h} = \gamma \mathit{B} \cdot \mathbf{S},
\end{equation}
\\
where $\gamma$ is the electron gyromagnetic ratio, $2.8$ MHz/G. After diagonalizing this matrix and including the spin-strain interaction Hamiltonian, the following equation is obtained:

\begin{equation}
    f_{\pm} = D + M_z \pm \sqrt{ (\gamma B_z)^2 + M_x^2 + M_y^2 }
    \label{f_pm_B}
\end{equation}
\\
where $f_{\pm}$ is the frequency of each spin transition. From Equation~\ref{f_pm_B}, the following relations can be derived:

\begin{equation}
    \Delta f_{\mathrm{shift}} = \frac{1}{2}(f_+ + f_-) - D = \mathit{M}_z
    \label{f_shift}
\end{equation}
\\
where $M_z$ is the volumetric (axial) strain amplitude, quantifying lattice compression or expansion along the NV center axis, directly affecting the resonance frequency shift observed in ODMR measurements, and:

\begin{equation}
    \Delta f_{\mathrm{split}} = f_+ - f_- = 2 \sqrt{ (\gamma B_z)^2 + M_x^2 + M_y^2 }
    \label{f_split}
\end{equation}
\\
where $M_x$ and $M_y$ represent the shear (transverse) strain components, corresponding to lattice distortions perpendicular to the NV axis. These transverse components primarily influence the magnitude of peak splitting in the ODMR spectra. Therefore, the effects of strain can be directly characterized by ODMR \textsuperscript{[30]}, as lattice distortions lead to measurable frequency shifts and splittings of the spin resonance transitions.


\section{Experimental Section}

\begin{figure}[ht]
    \centering
    \includegraphics[width=\linewidth]{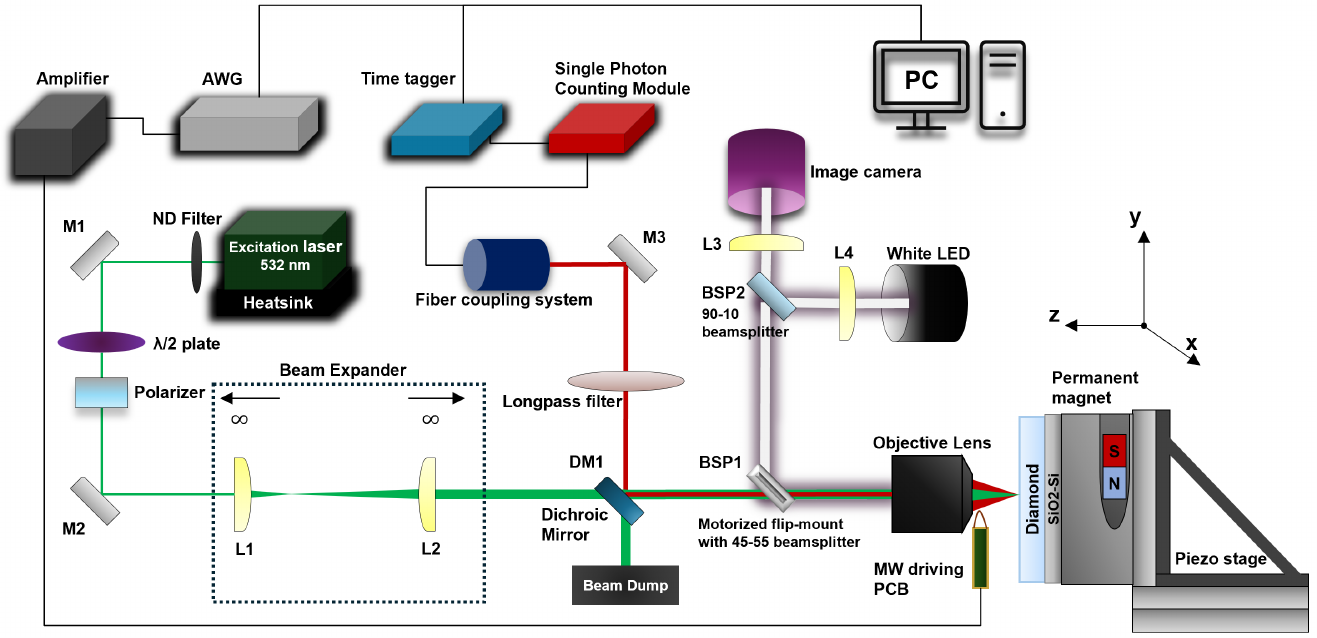}
    \caption{Schematic illustration of the confocal Optically Detected Magnetic Resonance (ODMR) setup used in this work. A 532 nm CW-laser provides optical excitation of the NV centers, with fluorescence collected via a high-NA objective and directed through a dichroic mirror and filters to a single-photon counting module (SPCM). Microwave excitation is delivered via a wire loop placed near the diamond surface. A bias magnetic field ($B_z$) is applied along the NV axis to lift spin degeneracy, enabling detection of frequency shifts and splittings associated with strain.}
    \label{fig:odmr-setup}
\end{figure}

The primary objective of this study is to measure the effects of strain in diamonds that have undergone direct bonding with an insulator ($\mathrm{SiO_2/Si}$) substrate. To achieve this, NV centers are used due to their excellent properties that make them highly sensitive to external magnetic fields and strain fields. As already described above, strain can cause splitting and shifts in the energy levels of NV centers. To enable these measurements, a confocal Optically Detected Magnetic Resonance (ODMR) setup is used, as seen in Figure~\ref{fig:odmr-setup}. With ODMR, these changes in the energy states and spins can be quantified in the form of the presence of magnetic, electric, strain, or temperature variations. The frequency shift $\Delta f_{\mathrm{shift}}$ and frequency splitting $\Delta f_{\mathrm{split}}$ are related to the strain tensor components $\epsilon_{\mathrm{ij}}$, through Equation~\ref{Full_M-N_equation}.

Solving Equation~\ref{f_shift} for  for $M_z$ gives:

\begin{equation}
    M_z = \Delta f_{\mathrm{shift}}
    \label{M_z-value}
\end{equation}
\\
while solving Equation~\ref{f_split} for $M_{xy} = \sqrt{M_x^2 + M_y^2}$ results in:

\begin{equation}
    M_{xy} = \sqrt{ \frac{\Delta f_{\mathrm{split}}^2}{4} - (\gamma B_z)^2 }
     \label{M_xy-value}
\end{equation}

\begin{figure}[ht]
    \includegraphics[width=\linewidth]{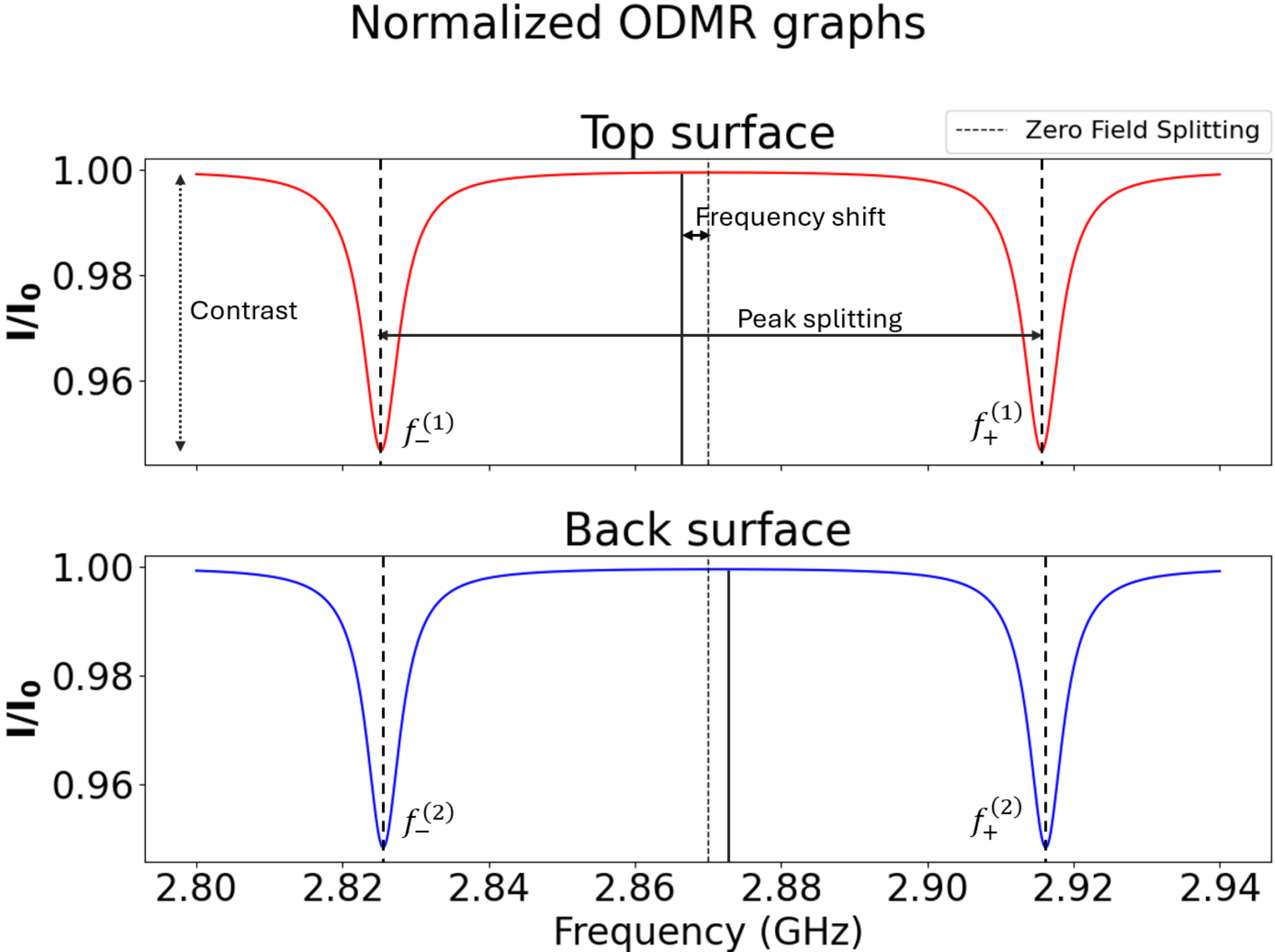}
    \caption{Indicative ODMR spectra obtained at the top and at the back surface.}
    \label{fig:ODMR_Graphs}
\end{figure}

 Therefore, by measuring the strain amplitudes ($M_x$, $M_y$, $M_z$) through ODMR measurements, one can partially reconstruct the strain tensor. Specifically, the volumetric strain component, $M_z$, provides a linear combination of the axial strain terms ($\epsilon_{xx}$, $\epsilon_{yy}$, and $\epsilon_{zz}$), weighted according to their orientation with respect to the NV center's crystallographic axis. In this research, the difference in values for $M_{xy}$, and $M_z$ is acquired, after comparing ODMR-obtained spectra at different depths of the diamond's volume, as illustrated in Figure~\ref{fig:ODMR_Graphs}.

For the experiment, a $10 \times 10\,\mathrm{mm^2}$ bulk diamond substrate (supplied by EDP), with a thickness of $\sim 300$ \textmu m is investigated, hereafter referred to as $\mathrm{DS_{PB}}$. It is a CVD-grown substrate, cut along the \{100\} crystallographic plane, that contains randomly distributed ensembles of NV centers with a concentration of roughly 15 ppb. The diamond substrate was double-side polished, exhibiting an initial surface roughness of $R_a < 2$ nm, as specified by the supplier. The SiO\textsubscript{2} surface on the Si wafer, prior to bonding, exhibited a root mean square (RMS) roughness of $S_q = 4.44$ nm, as measured by AFM.

The top (unbonded) surface of the diamond was later etched by approximately 50 \textmu m, using O$_2$ plasma, with no considerable change to surface roughness, in order to allow focusing on the back surface (diamond-$\mathrm{SiO_2/Si}$ interface). This challenge arose due to the objective lens' small working distance of 400 \textmu m and the 100 \textmu m diameter of the microwave delivery wire, which is placed between the objective and the piezo stage. Although the refractive index of diamond, $n = 2.4$, provides some leniency by modifying the effective distance of the objective-to-diamond, further etching of the surface ensures that proper focus on the back surface can be achieved and eliminates any worries of collision of the diamond with the objective during 2D scanning.

This diamond sample underwent the direct bonding process, but, due to its large surface area, the result was partial bonding of the DOI substrate, meaning that two regions of interest are present—both bonded and unbonded—on the same sample. Different strain fields may exist in these regions, since only part of its surface formed a strong bond with the insulator substrate underneath it. This bonded region has potentially experienced increased strain effects compared to the unbonded one. By comparing the ODMR peak splitting and frequency shift values at the top surface and at the bottom surface (diamond-$\mathrm{SiO_2/Si}$ interface), an overall increase in strain is observed in the interface.

It should be noted that the following measurements are performed in the presence of an external bias magnetic field $B_z=15.7$ G (or 43.9 MHz splitting) along the $z$-axis. A 532 nm green laser is used for optical excitation, at a power of 0.25 mW, and the emitted fluorescence is collected via a single-photon counting module (COUNT-10C-FC), coupled to a fiber.

\section{Results and Discussion}

\begin{figure}[h!]
    \centering
    \begin{picture}(0,0)
        \put(25,-7){\textbf{(a)}}
    \end{picture}
    \includegraphics[width=0.47\linewidth]{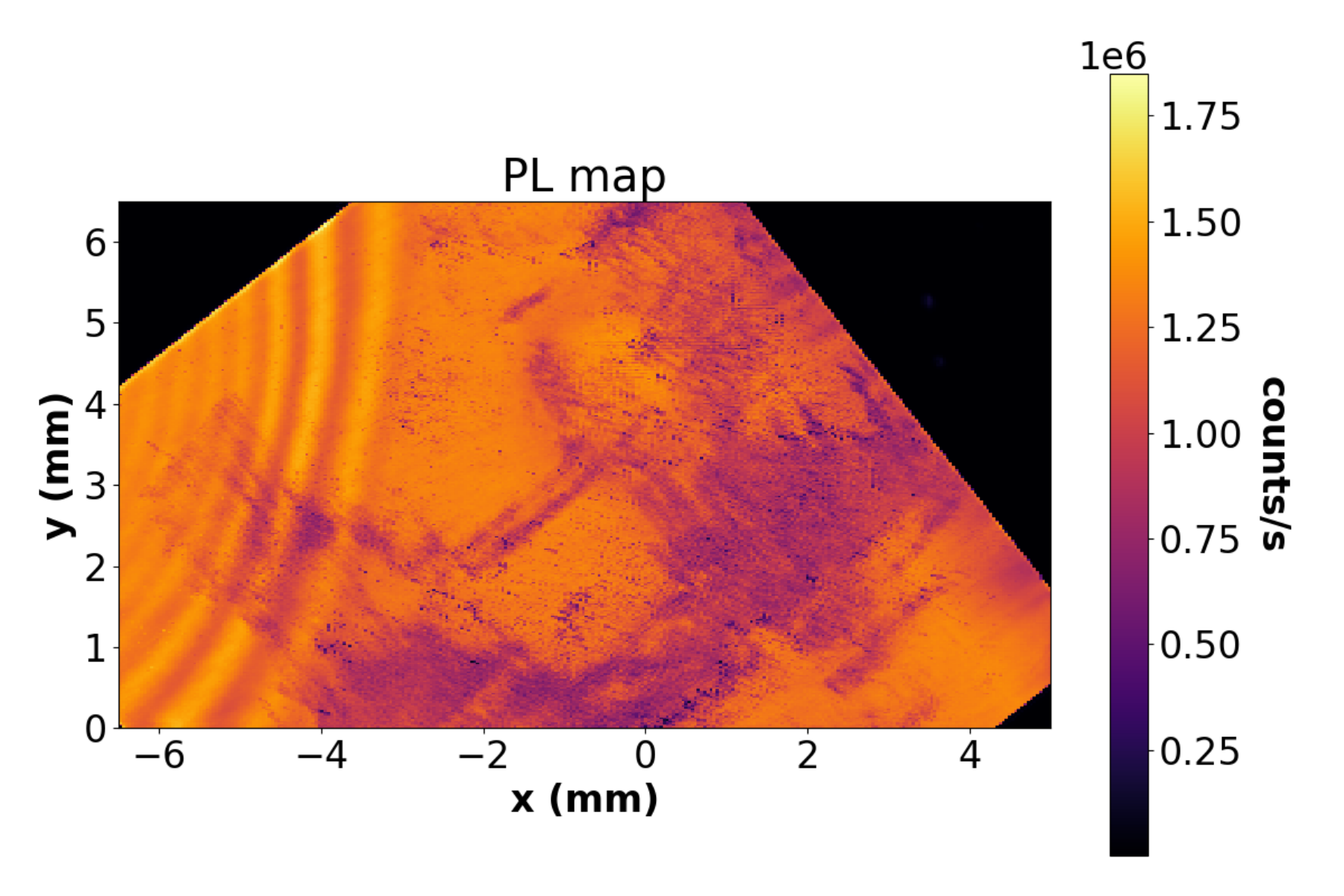}
    \hspace{0.5em}
    \begin{picture}(0,0)
        \put(25,-7){\textbf{(b)}}
    \end{picture}
    \includegraphics[width=0.47\linewidth]{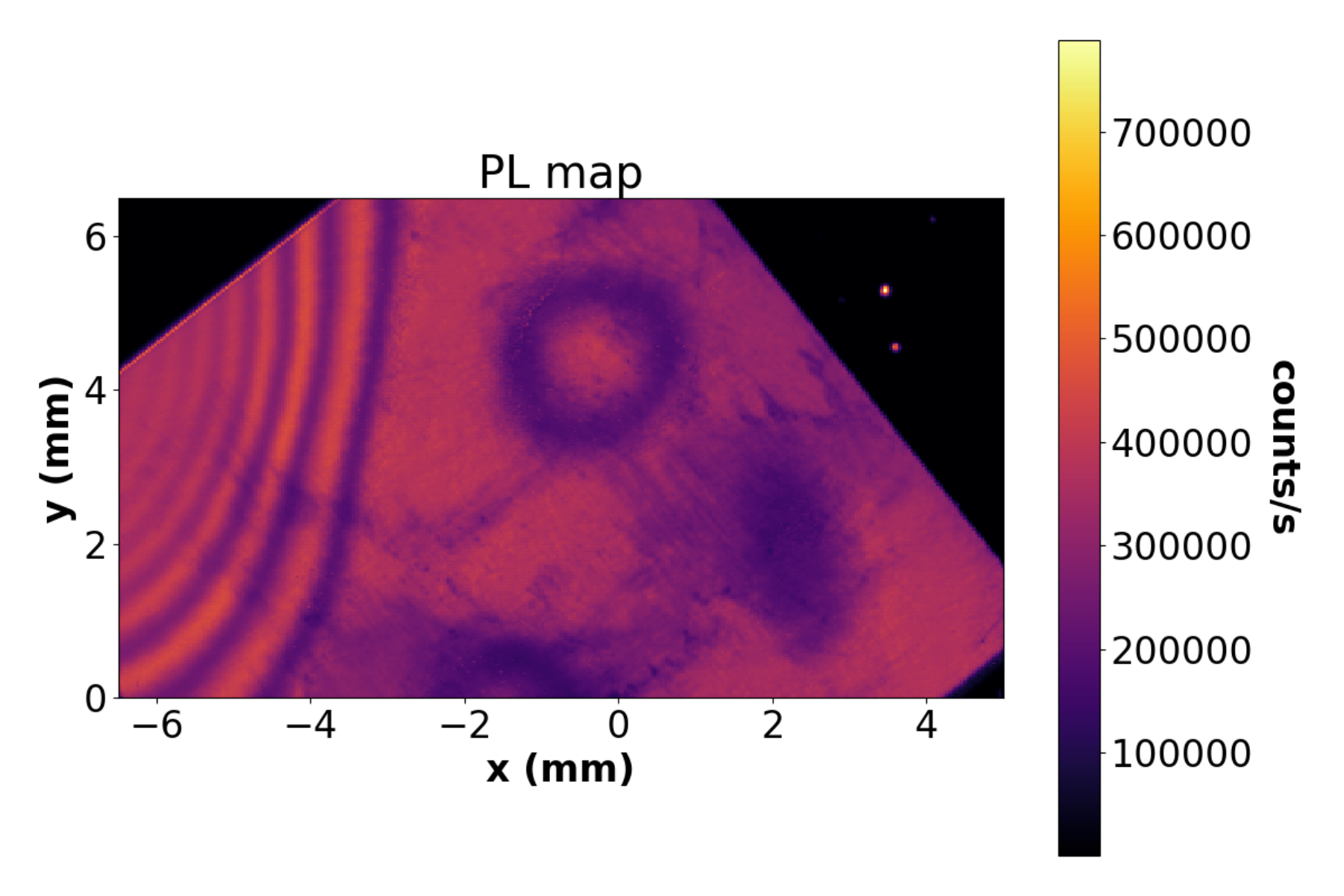}
    
    \vspace{2pt}
    
    \caption{Coarse PL maps of the partially-bonded diamond substrate.
    a) Top surface,  
    b) Back surface.}
    \label{fig:DS_PL}
\end{figure}

In Figure~\ref{fig:DS_PL}, differences in the acquired photoluminescence (PL) maps of the top and back surfaces are visible. During the direct bonding process, even with thorough cleaning of the substrates in a cleanroom environment, trapped particles or residues at the diamond-SiO\textsubscript{2}/Si interface may still remain, hindering proper bonding and inducing localized strain regions. This effect is particularly pronounced in the sample investigated here, owing to its large size of $10 \times 10\,\mathrm{mm^2}$. 

The bonded and unbonded regions can be clearly distinguished, providing a useful tool for evaluating the success and quality of the direct bonding process. Specifically, interference fringes are more apparent in the PL map closer to the bottom surface. This may indicate the presence of particles or impurities that were either not thoroughly rinsed or introduced during one of the direct bonding steps, and subsequently became trapped at the interface. Such particles could hinder proper adhesion and potentially lead to localized unbonded regions. These regions are linked to non-uniform bonding or strain effects that can alter the observed layer thickness and local refractive index. The direct bonding quality has to be improved, especially regarding substrates with larger surfaces. This would enable high-quality, reproducible bonding and more precise strain mapping.

\begin{figure}[h!]
    \centering
    \begin{picture}(0,0)
        \put(25,-7){\textbf{(a)}}
    \end{picture}
    \includegraphics[width=0.47\linewidth]{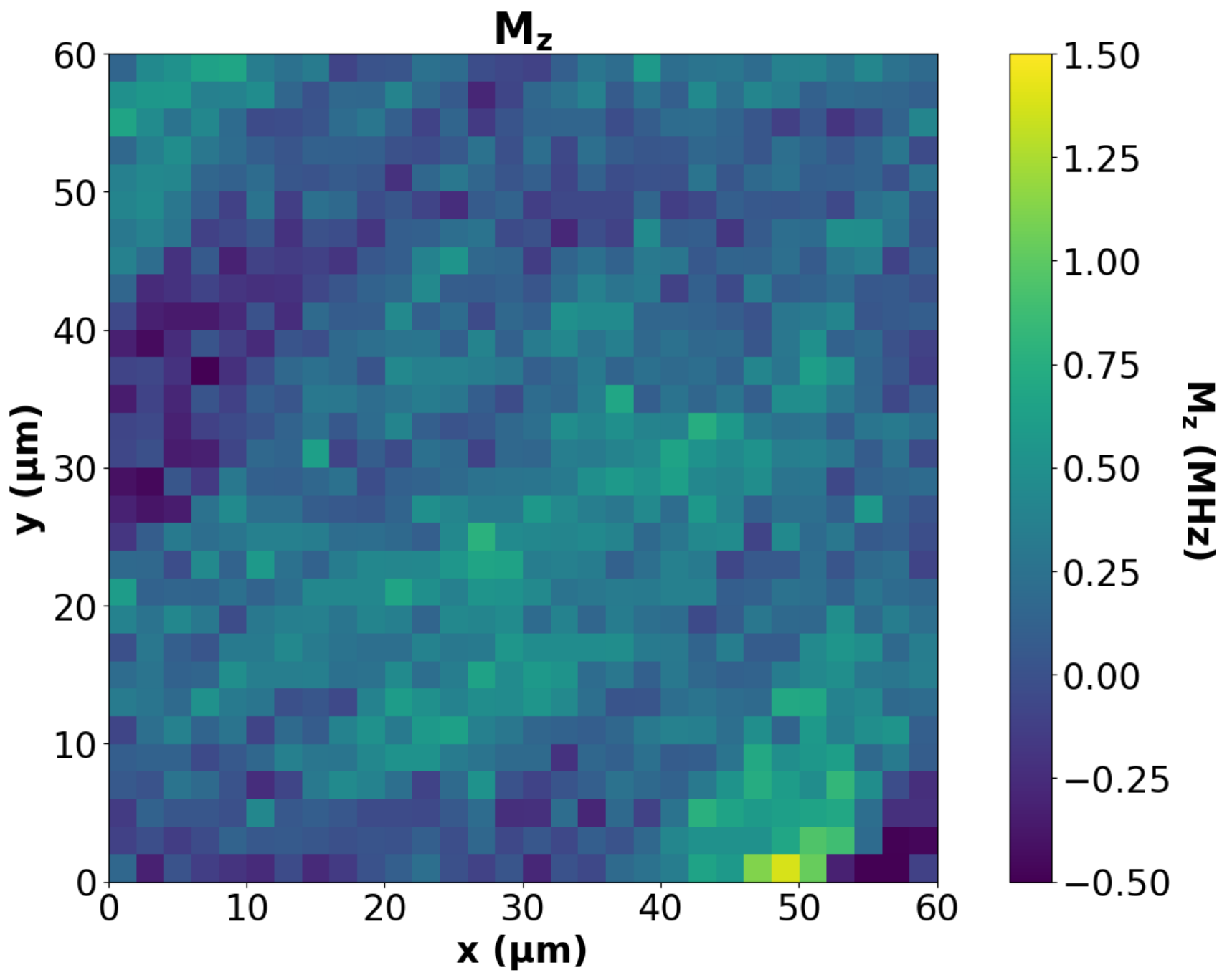}
    \hspace{0.5em}
    \begin{picture}(0,0)
        \put(25,-7){\textbf{(b)}}
    \end{picture}
    \includegraphics[width=0.47\linewidth]{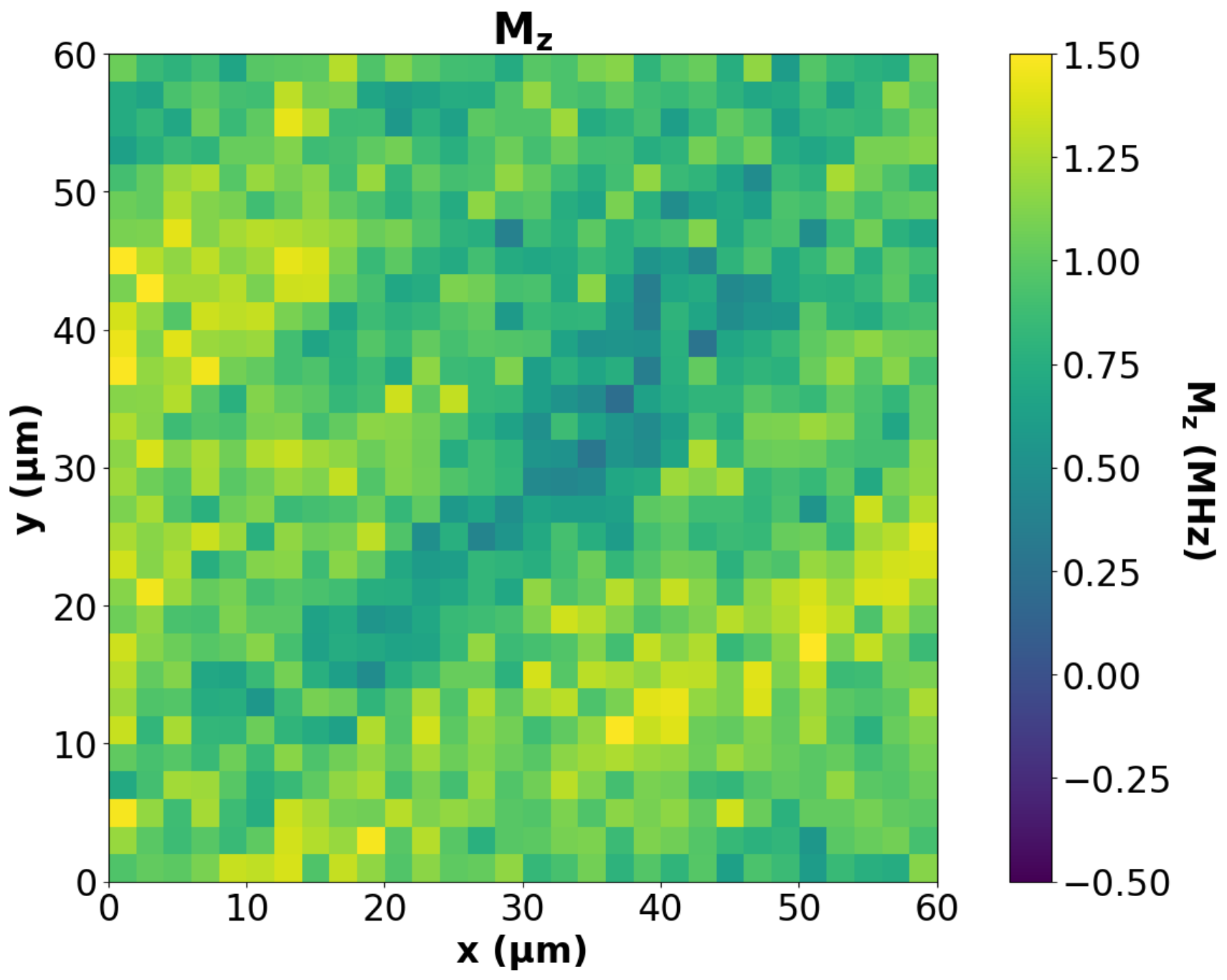}
    \vspace{2pt}
    \caption{ODMR 2D scans performed in a $60\times 60$ \textmu m$^2$ area, depicting the effects of the volumetric strain amplitude $M_z$.
    a) Top surface,  
    b) Back surface.}
    \label{fig:D2_FS}
\end{figure}

Afterward, 2D ODMR maps were obtained, starting with a $60\times 60$ \textmu m$^2$ area at the top surface, and then focusing on the diamond plane close to the diamond-$\mathrm{SiO_2/Si}$ interface. Figure~\ref{fig:D2_FS} represents volumetric strain variation, obtained by calculating the frequency shift $\Delta f_{\mathrm{shift}}$ of every point in the map, relative to the zero field splitting (ZFS) value of $D=$ 2.87 GHz, and solving for $M_z$, as seen in Equation~\ref{M_z-value}.

The milder volumetric strain effects at the top surface are attributed to intrinsic strain in the diamond. However, moving closer towards the back surface shows increased strain effects, which are assumed to have been caused by the thermal expansion mismatch that occurs during the direct bonding process.

\begin{figure}[h!]
    \centering
    \begin{picture}(0,0)
        \put(25,-7){\textbf{(a)}}
    \end{picture}
    \includegraphics[width=0.47\linewidth]{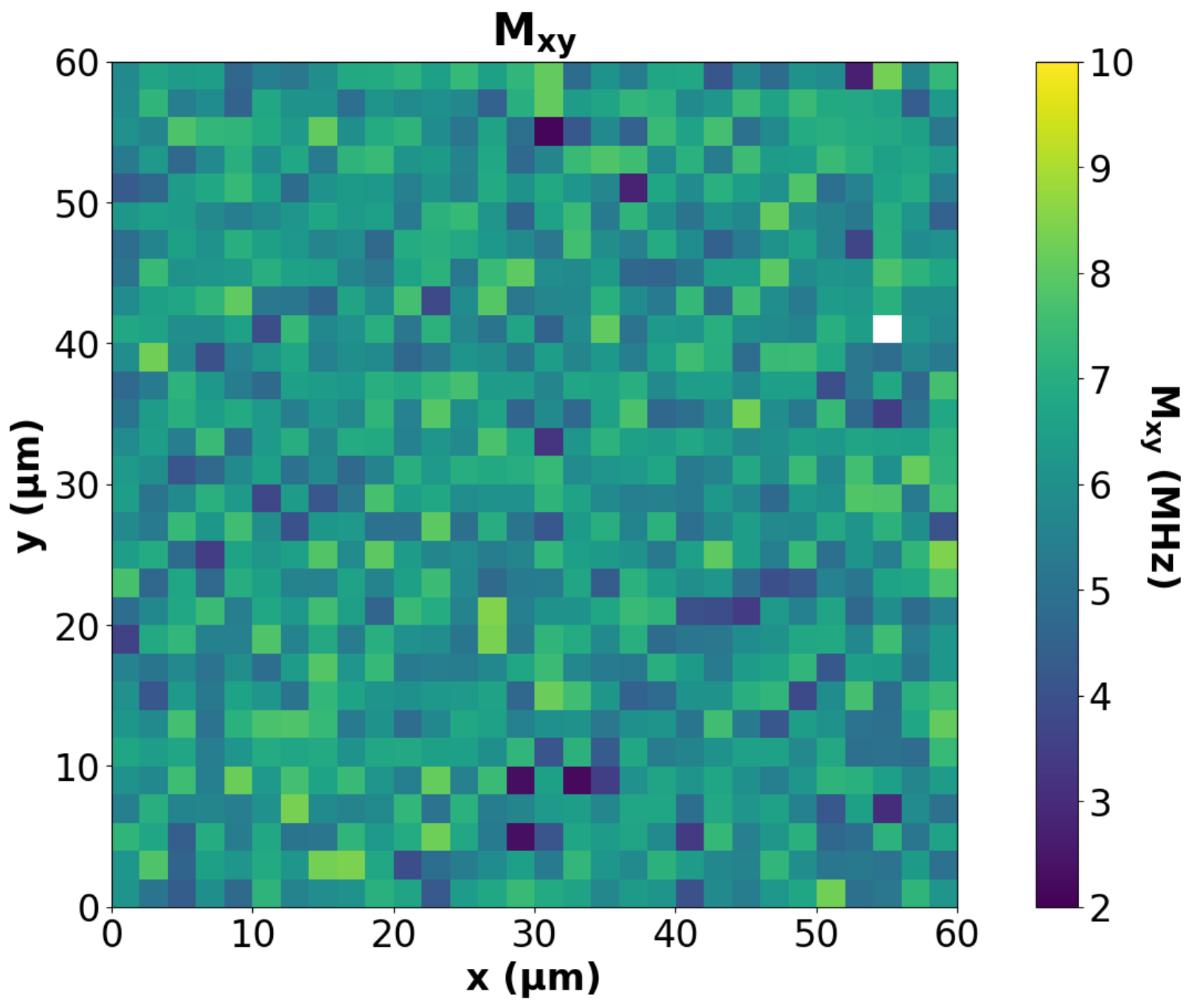}
    \hspace{0.5em}
    \begin{picture}(0,0)
        \put(25,-7){\textbf{(b)}}
    \end{picture}
    \includegraphics[width=0.47\linewidth]{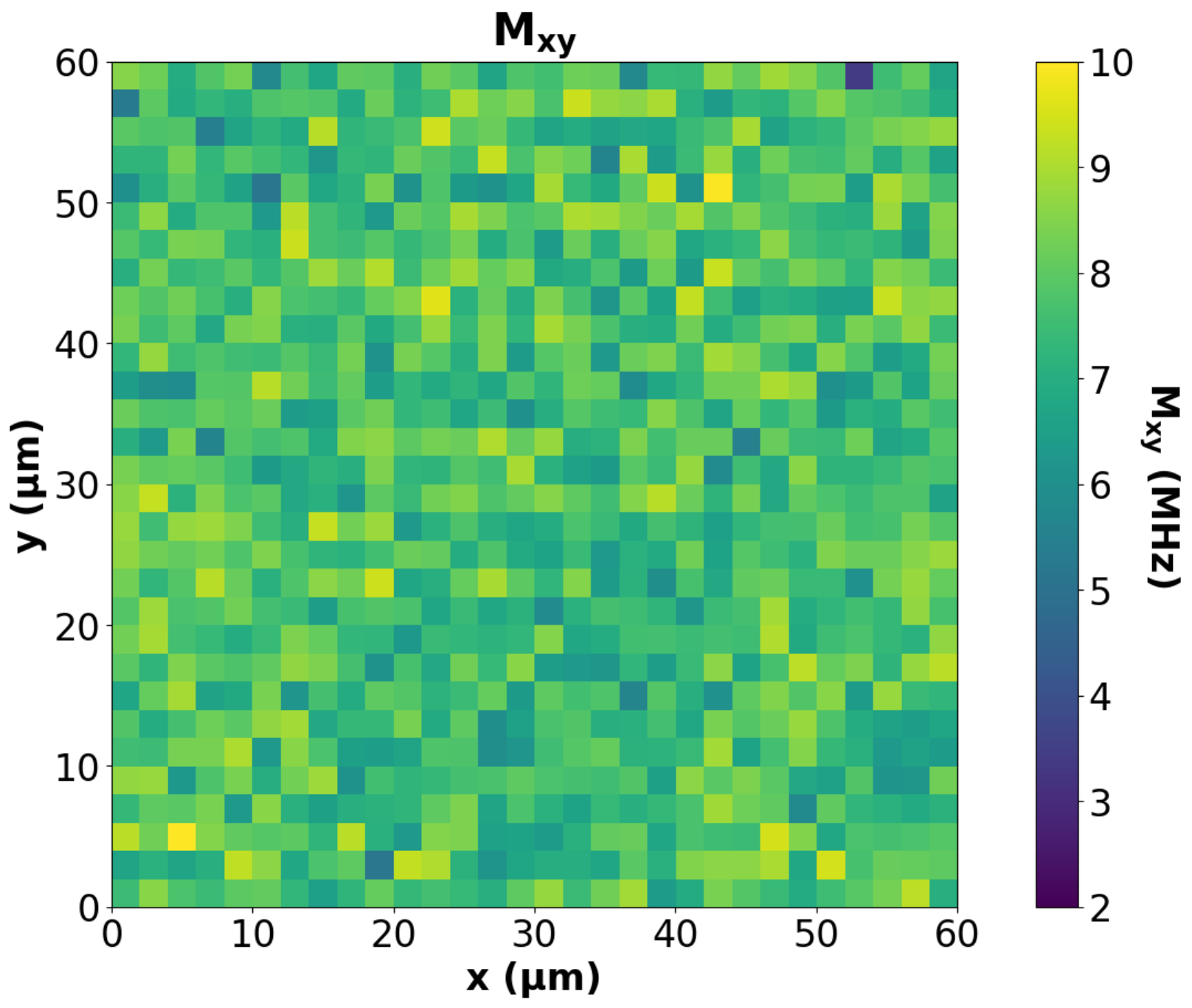}
    
    \vspace{2pt}
    
    \caption{ODMR 2D scans performed in a $60\times 60$ \textmu m$^2$ area, depicting the effects of the transverse strain amplitude $M_{xy}$.
    a) Top surface,  
    b) Back surface.}
    \label{fig:G1_PS}
\end{figure}

Similarly, Figure~\ref{fig:G1_PS} demonstrates the spatial variation of shear (transverse) strain, obtained via 2D ODMR mapping by analyzing the peak splitting at each measurement point according to Equation~\ref{M_xy-value}. Calculating the transverse strain amplitude ($M_{xy}$) reveals the magnitude of lattice distortions in planes perpendicular to the NV axis (i.e., the $xy$, $yz$, and $xz$ planes). An increase in shear strain is observed near the diamond–SiO\textsubscript{2}/Si interface, consistent with expectations based on the thermal expansion coefficient mismatch between the bonded materials.

Next, ODMR measurements were performed at various depths ($z$-axis scans), starting from the top surface of the $\mathrm{DS_{PB}}$ sample and finishing close to the bottom surface (diamond-SiO\textsubscript{2}/Si interface). These ODMR measurements provide information on multiple quantities, such as the frequency shift, which is related to volumetric strain, the frequency splitting, related to shear strain, ODMR peak contrast, which provides insight into the signal-to-noise ratio (SNR) and peak broadening. These measurements include data from both bonded and seemingly-unbonded regions.

\begin{figure}[h!]
    \centering
    \begin{picture}(0,0)
        \put(25,-7){\textbf{(a)}}
    \end{picture}
    \includegraphics[width=0.48\linewidth]{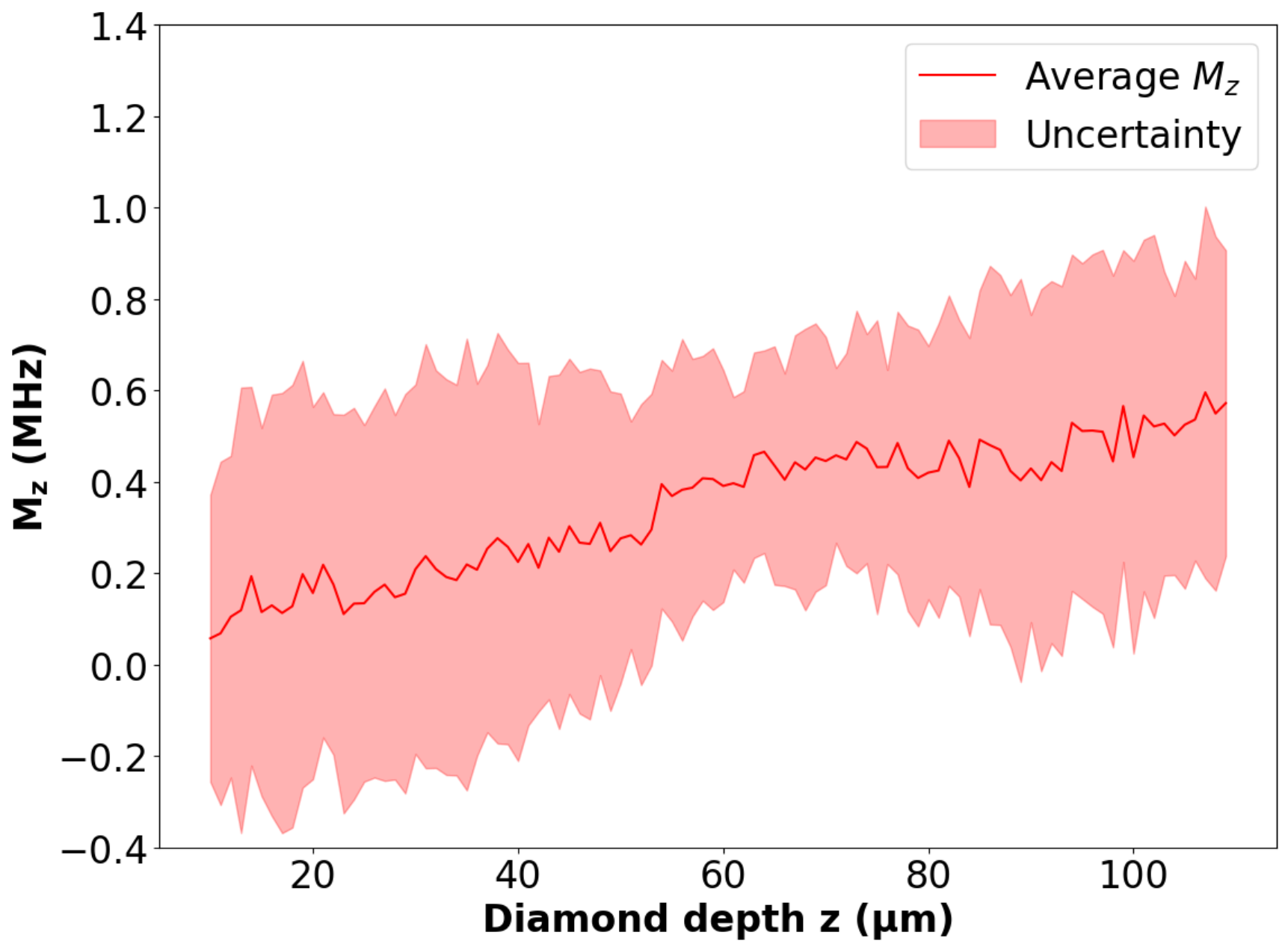}
    \hspace{0.5em}
    \begin{picture}(0,0)
        \put(25,-7){\textbf{(b)}}
    \end{picture}
    \includegraphics[width=0.48\linewidth]{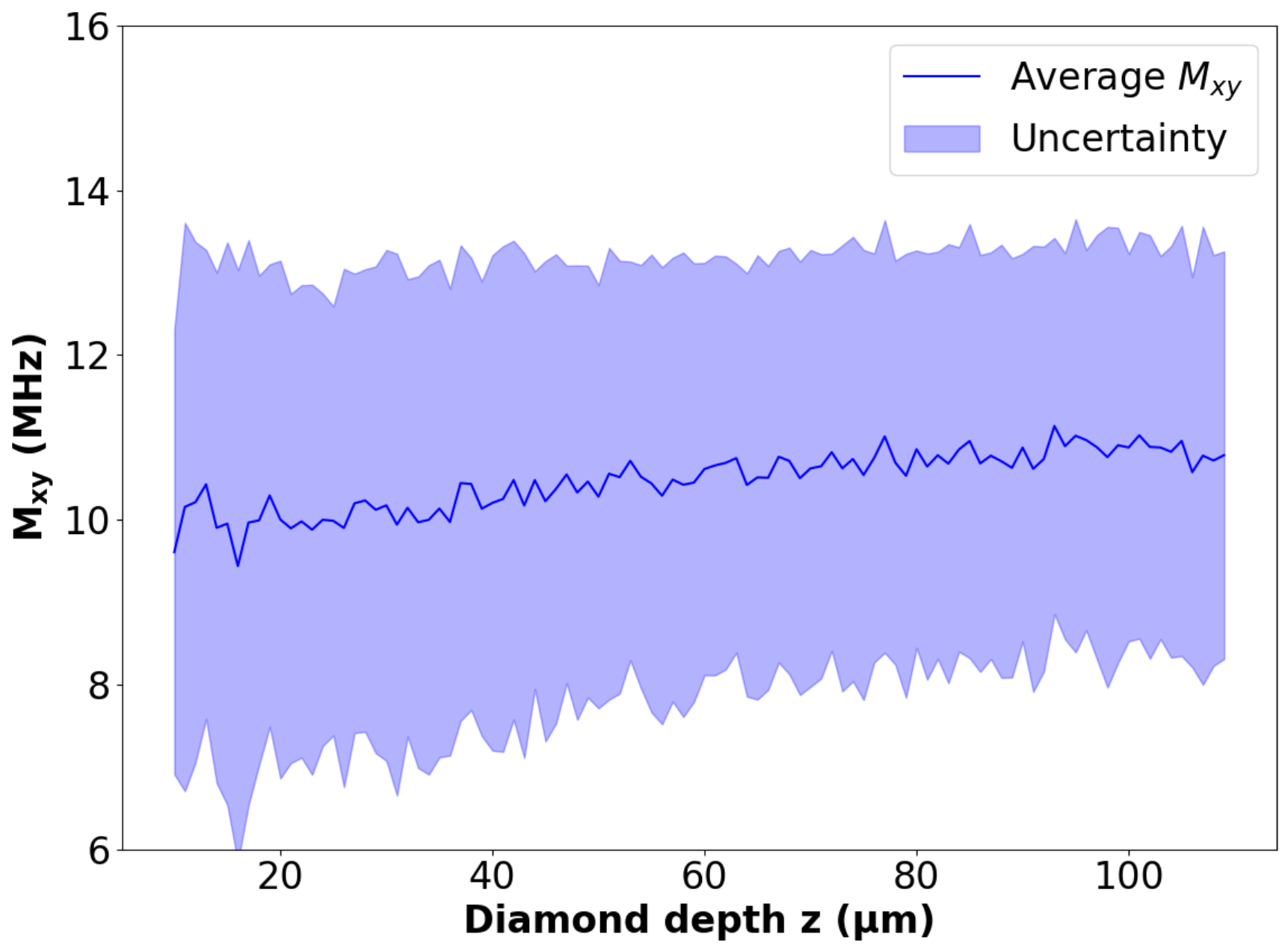}
    
    \vspace{2pt}
    
    \caption{Data from different ODMR graphs, obtained at varying planes of the diamond substrate. The data is averaged out of nine different ODMR scans, each starting from a different region of the surface of the diamond. a) Shows the increase of the $M_z$ parameter, acquired via frequency shift compared to the default ZFS value in ODMR measurements, and corresponding to volumetric strain, and b) shows the increase of the $M_{xy}$ parameter, acquired via peak splitting in ODMR measurements, and corresponding to shear strain components.}
    \label{fig:ZODMR_Mz_Mxy}
\end{figure}

In Figure~\ref{fig:ZODMR_Mz_Mxy}, a gradual increase in both volumetric and shear strain can be seen. Contrary to the 2D ODMR maps discussed earlier, such a measurement allows a direct observation of how the strain effects are increasingly more intense close to the interface. Values of $M_z$ range from 0.11 MHz up to 0.56 MHz, whereas values of $M_{xy}$ range from 10.05 MHz to 10.76 MHz.

\begin{figure}[h!]
    \centering
    \begin{picture}(0,0)
        \put(25,-7){\textbf{(a)}}
    \end{picture}
    \includegraphics[width=0.48\linewidth]{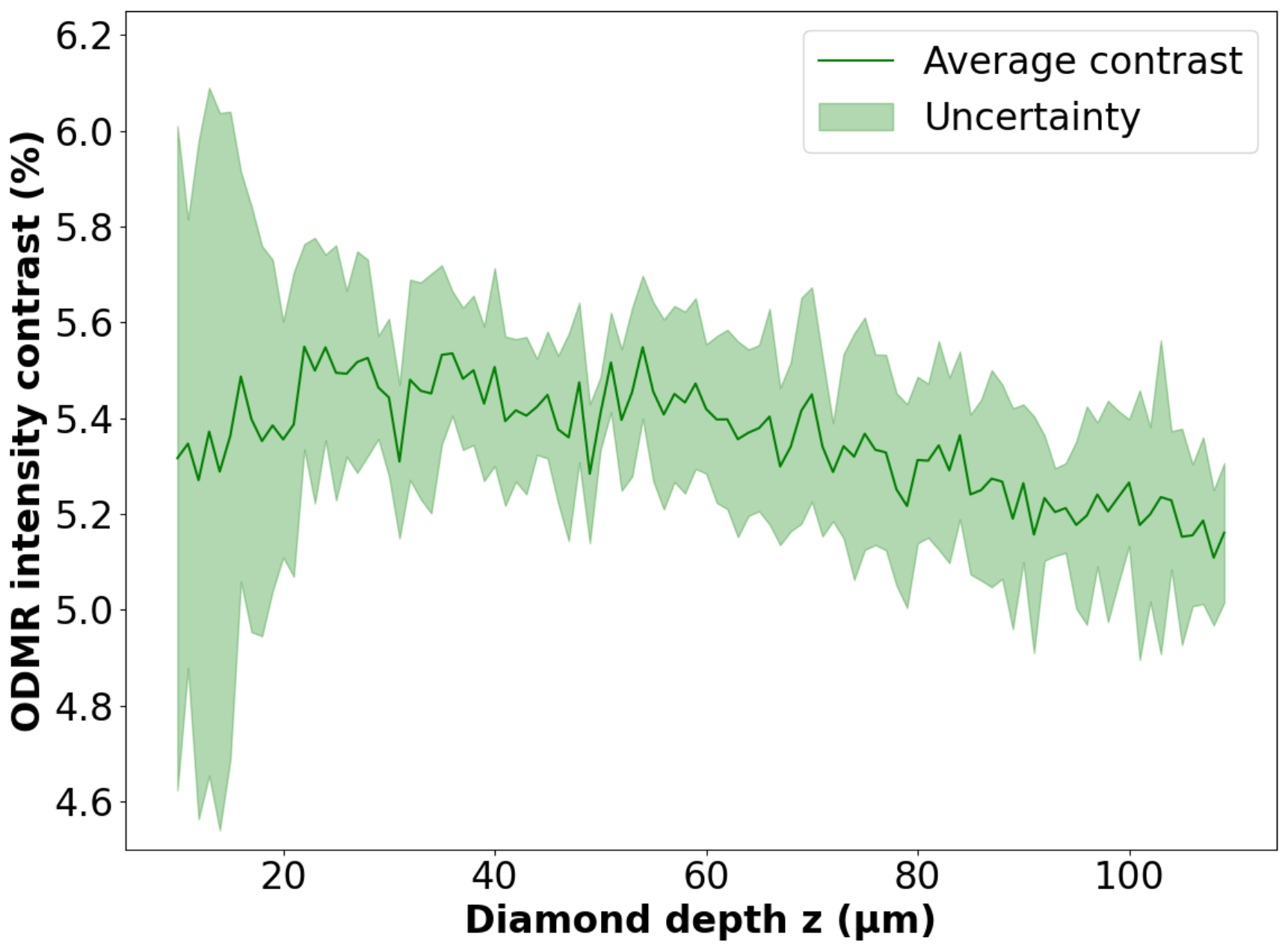}
    \hspace{0.5em}
    \begin{picture}(0,0)
        \put(25,-7){\textbf{(b)}} 
    \end{picture}
    \includegraphics[width=0.48\linewidth]{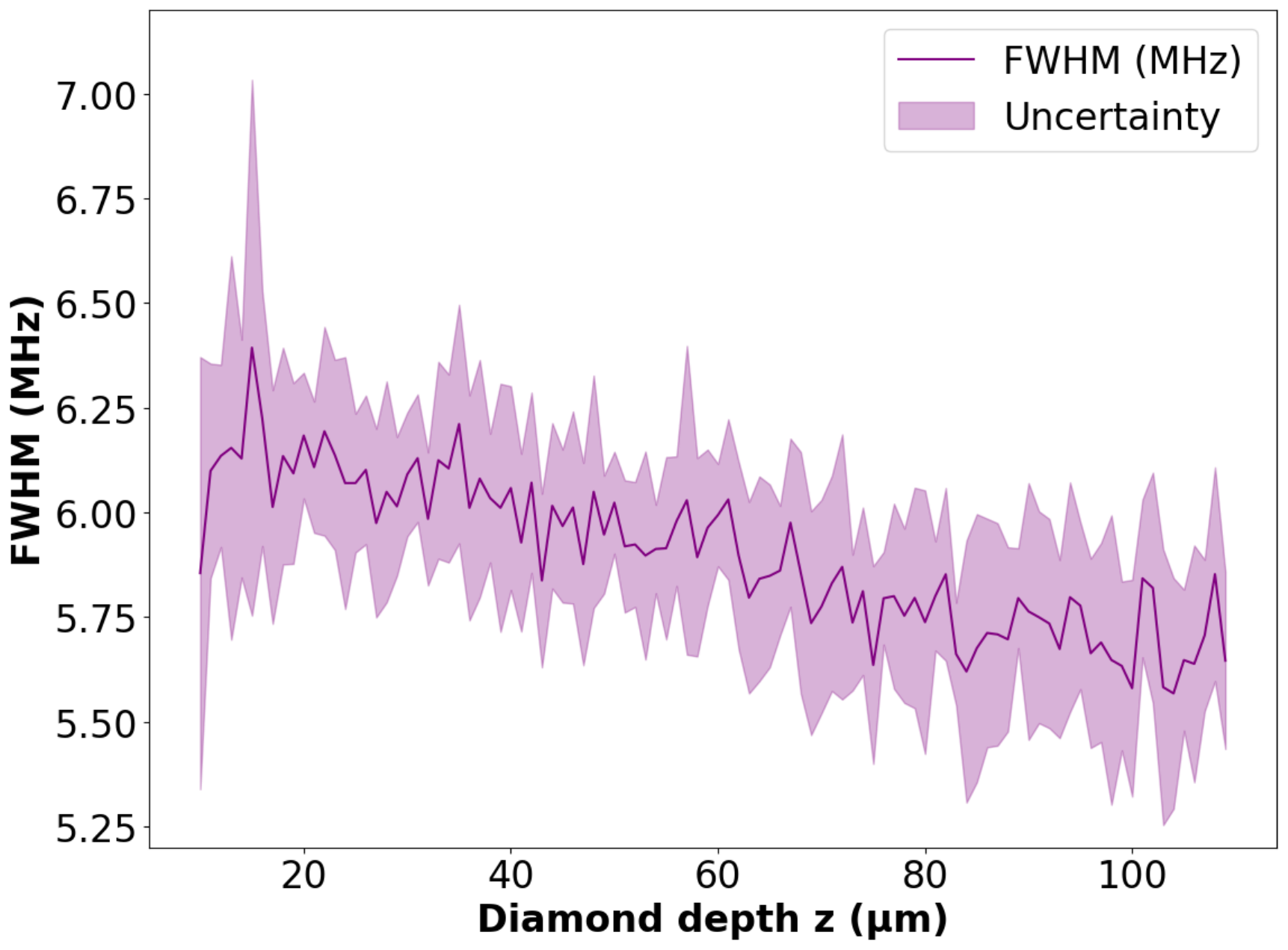}
    
    \vspace{2pt}
    
    \caption{Data from different ODMR graphs, obtained at varying planes of the diamond substrate. The data is averaged out of nine different ODMR scans, each starting from a different region of the surface of the diamond. \textbf{a)} Shows the decline in average ODMR intensity contrast. \textbf{b)} Shows the decline in the FWHM of the ODMR peaks.}
    \label{fig:ZODMR_Contrast_FWHM}
\end{figure}

The contrast of the peaks in an ODMR graph can provide important information on the quality of the color centers. It not only regards the signal-to-noise ratio (SNR), but is also affected by effects of decoherence and inhomogeneous broadening, which significantly impact the optical properties of the emitters. These phenomena can be caused by many factors, including the presence of strain fields which are investigated here. Figure~\ref{fig:ZODMR_Contrast_FWHM} demonstrates how contrast deteriorates close to the diamond-SiO\textsubscript{2}/Si interface. The initial ODMR intensity contrast at the top surface is approximately 5.32\%, increasing to an average of 5.51\% in the region where the NV centers exhibit the best signal, and finally decreasing to approximately 5.15\% close to the interface.

The average Full Width Half Maximum (FWHM) of the peaks in the ODMR spectra follows a similar trend, as it seems to diminish close to the interface. It drops from an average of $6.08$ MHz close to the top surface, down to an average of $5.70$ MHz close to the interface. This is an unexpected result, as one would anticipate that the observed linewidth worsens under increased strain conditions, similar to how contrast decreases.

Regarding the experimental setup, problems occasionally arise when focusing on the back surface of a thick substrate, due to limits set by the working distance of the objective lens, as well as obstruction by the microwave delivery wire. This necessitates thinning of the diamond layer by O\textsubscript{2} plasma etching; a rather time-consuming process. Lastly, the ODMR measurement process itself is another factor to take into consideration concerning the quality of the measurements. Long scanning durations make the samples more prone to thermal drifts and can introduce experimental parameter fluctuations. The time inefficiency is further exacerbated by the long direct bonding processing time of approximately a week. It is worth noting that future DOI substrates may be comprised of diamonds with a thickness of a few micrometers or even in the nanometer range. In such cases, the aforementioned strain effects and the quality of the emitters are bound to differ and be more localized.

\section{Conclusion}

Strain characterization was performed on a diamond directly bonded to an insulator substrate (SiO\textsubscript{2}/Si). The mismatch in thermal expansion coefficients between diamond and SiO\textsubscript{2}/Si during the direct bonding process induces strain effects, particularly pronounced near the interface. In this study, strain properties were evaluated across regions exhibiting different bonding conditions by conducting depth-resolved measurements using optically detected magnetic resonance (ODMR). The ODMR technique allowed for the correlation of resonance frequency shifts and splittings to volumetric and shear strain amplitudes, respectively. Results revealed increased lattice deformations when approaching the diamond–SiO\textsubscript{2}/Si interface, with the volumetric strain parameter exhibiting an increase of 0.45 MHz and the shear strain parameter rising by 0.71 MHz. Additionally, photoluminescence (PL) mapping of partially bonded diamond areas identified interference fringes that likely indicate localized strain or imperfect bonding at the interface. In this investigation, only minor changes in ODMR contrast and linewidth were observed, suggesting negligible deterioration in emitter quality. Specifically, contrast deteriorated by approximately 0.36\%, while unexpectedly, the linewidth (FWHM) also decreased by 0.38 MHz. In conclusion, we have determined that PL mapping serves as an effective preliminary method for assessing bonding quality, while ODMR provides detailed insights into the magnitude and nature of strain and its impact on NV centers. Nevertheless, further detailed measurements are required to fully understand how strain influences the optical and electronic properties of color centers in directly bonded diamond substrates.

\medskip
\textbf{Conflict of Interest}
The authors declare no conflict of interest.

\medskip
\textbf{Data Availability Statement}
The data that support the findings of this study are available from the corresponding authors upon reasonable request.

\medskip
\textbf{Acknowledgements}
We gratefully acknowledge support from the joint research program “Modular quantum computers” by Fujitsu Limited and Delft University of Technology, co-funded by the Netherlands Enterprise Agency under project number PPS2007.

\medskip

%

\textbf{References}\\

[1] A. Faraon, C. Santori, Z. Huang, V. M. Acosta, R. G. Beausoleil, Coupling of Nitrogen-Vacancy Centers to Photonic Crystal Cavities in Monocrystalline Diamond, \textit{Phys. Rev. Lett.} \textbf{2012}, \textit{109}, 033604, https://doi.org/10.1103/PhysRevLett.109.033604\\

[2] J.-C. Arnault, S. Saada, V. Ralchenko, Chemical Vapor Deposition Single-Crystal Diamond: A Review, \textit{Phys. Status Solidi RRL} \textbf{2022}, \textit{16}, 2100354, https://doi.org/10.1002/pssr.202100354\\

[3] S. W. Ding, M. Haas, X. Guo, K. Kuruma, C. Jin, Z. Li, D. D. Awschalom, N. Delegan, F. J. Heremans, A. High, M. Lončar, High-Q Cavity Interface for Color Centers in Thin Film Diamond, \textit{arXiv} \textbf{2024}, arXiv:2402.05811, https://arxiv.org/abs/2402.05811\\

[4] R. Ishihara, J. Hermias, S. Yu, K. Y. Yu, Y. Li, S. Nur, T. Iwai, T. Miyatake, K. Kawaguchi, Y. Doi, S. Sato, 3D integration technology for quantum computer based on diamond spin qubits, \textit{Proc. IEEE Int. Electron Devices Meet. (IEDM)} \textbf{2021}, \textit{14.5.1–14.5.4}, https://doi.org/10.1109/IEDM19574.2021.9720552\\

[5] P. E. Barclay, K.-M. C. Fu, C. Santori, A. Faraon, R. G. Beausoleil, Hybrid Nanocavity Resonant Enhancement of Color Center Emission in Diamond, \textit{Phys. Rev. X} \textbf{2011}, \textit{1}, 011007, https://doi.org/10.1103/PhysRevX.1.011007\\

[6] M. Ruf, N. H. Wan, H. Choi, D. Englund, R. Hanson, Quantum networks based on color centers in diamond, \textit{J. Appl. Phys.} \textbf{2021}, \textit{130}, 070901, https://doi.org/10.1063/5.0056534\\

[7] K. Ngan, Y. Zhan, C. Dory, J. Vučković, S. Sun, Quantum Photonic Circuits Integrated with Color Centers in Designer Nanodiamonds, \textit{Nano Lett.} \textbf{2023}, \textit{23}, 9360--9366, https://doi.org/10.1021/acs.nanolett.3c02645\\

[8] S. Mi, M. Kiss, T. Graziosi, N. Quack, Integrated photonic devices in single crystal diamond, \textit{J. Phys. Photonics} \textbf{2020}, \textit{2}, 042001, https://doi.org/10.1088/2515-7647/aba171\\

[9] F. Lenzini, N. Gruhler, N. Walter, W. H. P. Pernice, Diamond as a Platform for Integrated Quantum Photonics, \textit{Adv. Quantum Technol.} \textbf{2018}, \textit{1}, 1800061, https://doi.org/10.1002/qute.201800061\\

[10] S. Sun, J. L. Zhang, K. A. Fischer, M. J. Burek, C. Dory, K. G. Lagoudakis, Y.-K. Tzeng, M. Radulaski, Y. Kelaita, A. Safavi-Naeini, Z.-X. Shen, N. A. Melosh, S. Chu, M. Lončar, J. Vučković, Cavity-Enhanced Raman Emission from a Single Color Center in a Solid, \textit{Phys. Rev. Lett.} \textbf{2018}, \textit{121}, 083601, https://doi.org/10.1103/PhysRevLett.121.083601\\

[11] T. Chen, J. Hermias, S. Nur, R. Ishihara, Hydrophilic direct bonding of (100) diamond and deposited SiO\textsubscript{2} substrates, \textit{Appl. Phys. Lett.} \textbf{2025}, \textit{126}, 231901, https://doi.org/10.1063/5.0263008\\

[12] V. Masteika, J. Kowal, N. St. J. Braithwaite, T. Rogers, A Review of Hydrophilic Silicon Wafer Bonding, \textit{ECS J. Solid State Sci. Technol.} \textbf{2014}, \textit{3}, Q42, https://doi.org/10.1149/2.007403jss\\

[13] T. Matsumae, Y. Kurashima, H. Umezawa, H. Takagi, Hydrophilic direct bonding of diamond (111) substrate using treatment with H\textsubscript{2}SO\textsubscript{4}/H\textsubscript{2}O\textsubscript{2}, \textit{Jpn. J. Appl. Phys.} \textbf{2019}, \textit{59}, SBBA01, https://doi.org/10.7567/1347-4065/ab4c87\\

[14] T. Plakhotnik, M. W. Doherty, J. H. Cole, R. Chapman, N. B. Manson, All-Optical Thermometry and Thermal Properties of the Optically Detected Spin Resonances of the NV\textsuperscript{--} Center in Nanodiamond, \textit{Nano Lett.} \textbf{2014}, \textit{14}, 4989--4996, https://doi.org/10.1021/nl501841d\\

[15] A. Berzins, J. Smits, A. Petruhins, R. Rimsa, G. Mozolevskis, M. Zubkins, I. Fescenko, NV microscopy of thermally controlled stresses caused by thin Cr\textsubscript{2}O\textsubscript{3} films, \textit{Opt. Express} \textbf{2023}, \textit{31}, 17950--17963, https://doi.org/10.1364/OE.489901\\

[16] E. Oliveira, C. Li, X. Zhang, et al., Stability of oxygenated groups on pristine and defective diamond surfaces, \textit{MRS Adv.} \textbf{2022}, \textit{7}, 543--546, https://doi.org/10.1557/s43580-022-00242-1\\

[17] T. Matsumae, Y. Kurashima, H. Takagi, H. Umezawa, E. Higurashi, Low-temperature direct bonding of diamond (100) substrate on Si wafer under atmospheric conditions, \textit{Scr. Mater.} \textbf{2021}, \textit{191}, 52--55, https://doi.org/10.1016/j.scriptamat.2020.09.006\\

[18] C. Malhaire, M. Granata, D. Hofman, A. Amato, V. Martinez, G. Cagnoli, A. Lemaitre, N. Shcheblanov, Determination of stress in thin films using micro-machined buckled membranes, \textit{J. Vac. Sci. Technol. A} \textbf{2023}, \textit{41}, 043401, https://doi.org/10.1116/6.0002590\\

[19] I. Vlasov, V. Ralchenko, D. Zakharov, N. Zakharov, Intrinsic Stress Origin in High Quality CVD Diamond Films, \textit{Phys. Status Solidi A} \textbf{1999}, \textit{174}, 11--18, https://doi.org/10.1002/(SICI)1521-396X(199907)174:1<11::AID-PSSA11>3.0.CO;2-T\\

[20] S. Stoupin, Y. V. Shvyd'ko, Ultraprecise studies of the thermal expansion coefficient of diamond using backscattering x-ray diffraction, \textit{Phys. Rev. B} \textbf{2011}, \textit{83}, 104102, https://doi.org/10.1103/PhysRevB.83.104102\\

[21] F. Jansen, M. A. Machonkin, N. Palmieri, D. Kuhman, Thermal expansion and elastic properties of plasma-deposited amorphous silicon and silicon oxide films, \textit{Appl. Phys. Lett.} \textbf{1987}, \textit{50}, 1059--1061, https://doi.org/10.1063/1.97969\\

[22] H. Tada, A. E. Kumpel, R. E. Lathrop, J. B. Slanina, P. Nieva, P. Zavracky, I. N. Miaoulis, P. Y. Wong, Thermal expansion coefficient of polycrystalline silicon and silicon dioxide thin films at high temperatures, \textit{J. Appl. Phys.} \textbf{2000}, \textit{87}, 4189--4193, https://doi.org/10.1063/1.373050\\

[23] K. G. Lyon, G. L. Salinger, C. A. Swenson, G. K. White, Linear thermal expansion measurements on silicon from 6 to 340 K, \textit{J. Appl. Phys.} \textbf{1977}, \textit{48}, 865--868, https://doi.org/10.1063/1.323747\\

[24] C. A. Swenson, Recommended values for the thermal expansivity of silicon from 0 to 1000 K, \textit{J. Phys. Chem. Ref. Data} \textbf{1983}, \textit{12}, 179--182, https://doi.org/10.1063/1.555681\\

[25] J. T. L. Gamler, A. Leonardi, X. Sang, K. M. Koczkur, R. R. Unocic, M. Engel, S. E. Skrabalak, Effect of lattice mismatch and shell thickness on strain in core@shell nanocrystals, \textit{Nanoscale Adv.} \textbf{2020}, \textit{2}, 1105--1114, https://doi.org/10.1039/D0NA00061B\\

[26] R. H. Doremus, Oxidation of Silicon: Stress Relaxation in Silica, \textit{J. Electrochem. Soc.} \textbf{1987}, \textit{134}, 2001, https://doi.org/10.1149/1.2100806\\

[27] D. A. Broadway, B. C. Johnson, M. S. J. Barson, S. E. Lillie, N. Dontschuk, D. J. McCloskey, A. Tsai, T. Teraji, D. A. Simpson, A. Stacey, J. C. McCallum, J. E. Bradby, M. W. Doherty, L. C. L. Hollenberg, J.-P. Tetienne, Microscopic Imaging of the Stress Tensor in Diamond Using in Situ Quantum Sensors, \textit{Nano Lett.} \textbf{2019}, \textit{19}, 4543--4550, https://doi.org/10.1021/acs.nanolett.9b01402\\

[28] J. H. N. Loubser, J. A. van Wyk, Electron spin resonance in the study of diamond, \textit{Rep. Prog. Phys.} \textbf{1978}, \textit{41}, 1201, https://doi.org/10.1088/0034-4885/41/8/002\\

[29] P. Udvarhelyi, V. O. Shkolnikov, A. Gali, G. Burkard, A. Pályi, Spin-strain interaction in nitrogen-vacancy centers in diamond, \textit{Phys. Rev. B} \textbf{2018}, \textit{98}, 075201, https://doi.org/10.1103/PhysRevB.98.075201\\

[30] M. S. Alam, F. Gorrini, M. Gawełczyk, D. Wigger, G. Coccia, Y. Guo, S. Shahbazi, V. Bharadwaj, A. Kubanek, R. Ramponi, P. E. Barclay, A. J. Bennett, J. P. Hadden, A. Bifone, S. M. Eaton, P. Machnikowski, Determining strain components in a diamond waveguide from zero-field optically detected magnetic resonance spectra of negatively charged nitrogen-vacancy-center ensembles, \textit{Phys. Rev. Appl.} \textbf{2024}, \textit{22}, 024055, https://doi.org/10.1103/PhysRevApplied.22.024055\\





\end{document}